\documentclass[twocolumn,aps,showpacs]{revtex4}
\usepackage{epsfig}

\def\lsim{\:\raisebox{-0.5ex}{$\stackrel{\textstyle<}{\sim}$}\:}
\def\gsim{\:\raisebox{-0.5ex}{$\stackrel{\textstyle>}{\sim}$}\:}
\def\pb {{\bf p}}
\def\rf#1{{(\ref{#1})}}

\def\u#1{u(#1)}
\def\v#1{v(#1)}

\def\qb {{\bf q}}

\def\gmu{\gamma_{\mu}}

\def\gmmunnu{g^{\mu\nu}}

\def\smunu{\sigma_{\mu\nu}}

\def\kb {{\bf k}}
\def\gmmunnu{g^{\mu\nu}}
\def\be{\begin{equation}}
\def\ee{\end{equation}}
\def\br{\begin{eqnarray}}
\def\er{\end{eqnarray}}
\def\ket#1{|#1 \rangle}
\def\bra#1{\langle #1|}
\def\N{\cal N}
\def\bc{\begin{center}}
\def\ec{\end{center}}
\def\psix{\psi(x)}
\def\psixb{\bar{\psi}(x)}
\begin{document}
%\textwidth 7 inch
%\baselineskip=21pt

\title{ A new Fermi smearing approach for scattering of multi-GeV
electrons by nuclei}

\author{A. Mariano}
\affiliation{ Departamento de F\'\i sica, Facultad de Ciencias
Exactas, Universidad Nacional de la Plata, cc. 67, 1900 La
Plata, Argentina
\\ E-mail: mariano@venus.fisica.unlp.edu.ar}

\author{P. Podesta Lerma}
\affiliation{  Departamento F\'\i sica, Centro de Investigaci\'on y
de Estudios Avanzados del IPN,\\ Apdo. Postal 14-740, 07000
M\'exico, D.F., M\'exico
\\ E-mail: podesta@fnal.gov}

\begin{abstract}
The cross section for electron scattering by nuclei at high
momentum transfers is calculated within the Fermi smearing approximation (FSA),
where binding effects on the struck nucleon
are introduced via the relativistic Hartree approximation
(RHA). The model naturally preserves  current conservation, since
the response tensor for an off-shell nucleon conserves the same form that for
a free one but with an effective mass. Different parameterizations for
the inelastic nucleon structure function, are analyzed. The
smearing at the Fermi surface is introduced through a momentum distribution obtained
from a perturbative  nuclear matter calculation. Recent CEBAF
data on inclusive scattering of $4.05$ GeV electrons on
$^{56}$Fe are well reproduced for all measured geometries for the first time,
as is evident from the comparison with previous calculations.
\end{abstract}

\maketitle

\section{Introduction}
Electron scattering by nuclei at high momentum transfers is  a
powerful tool to study the effective constituents of hadronic
matter and exhibits  very interesting new features. Within the
$Q^{2}\equiv-q^{2} > 1(GeV/c)^2$ (or ${\rm q} \equiv |\qb|> 1
GeV/c$) domain, where $q \equiv ( \omega,{\bf q})$ is the
four-momentum transferred by the virtual photon,  the struck
nucleon is relativistic having a momenta of the order its mass
$M$. In addition for such  regime the probability for exciting
internal degrees of freedom of the nucleon (nucleon inelastic
response) becomes increasingly important. Ideally, to describe the
target response,  one should start from a relativistic covariant
theory of nuclei. However, such an approach is not practicable due
to the difficulties in treating the meson exchange interactions.
On the other hand, how to describe  a nucleon with three momentum
$\pb$ inside the nucleus  is a well known nonrelativistic nuclear
structure problem. Thus, a model that couples  both regimes is
necessary. Electron scattering experiments have been described
with a great variety of approximations, starting with the plane wave impulse approximation
PWIA. In the Born approximation the $A(e,e')A'$ differential cross
section reads
\bigskip
\be
\frac{d^{2}\sigma}{d\Omega'd\epsilon'}= \frac{e^{2}}{{\rm q}^{4}}
\frac{{\rm k'} }{{\rm k}}L^{\mu\nu}(k,k')
W^{A}_{\mu\nu}(\omega,{\bf q}),\label{1} \ee being ${\rm
k,k'}\equiv |\kb|, |\kb'|$, $W^{A}_{\mu\nu}$ the nuclear response
tensor, $L^{\mu\nu}(k,k')= 1/2[ k'^\mu k^\nu + k^\mu k'^\nu
+(q^2/2-m^2/2)\gmmunnu ]$ the lepton tensor describing incoming
and outgoing plane-wave electron states of four-momentum
$k=(\epsilon \equiv \sqrt{\kb '^2+m^2},\kb)$ and
$k'=(\epsilon'\equiv \sqrt{\kb '^2+m^2},\kb ')$ respectively, and
$\Omega ' \equiv (\theta, \phi)$ the scattering angle. The  PWIA
lies on the following assumptions:

i)the nuclear current operator can be written as the sum of the
one-body nucleon currents;

ii)the target decays virtually  into a on-shell (A-1) nucleus
(spectator) and the off-shell ($p^2 \neq M^2$) struck nucleon, of
four-momentum $p=(p_0,\pb)$ ; and

iii)the nucleon  that absorbes the photon is the same that
leaves the target without interaction with the spectator,
the final state interactions (FSI) being dropped.
Under these suppositions,  the nuclear response can be expressed as a convolution
\cite{Cio91}

\be
W^{A}_{\mu\nu}(\omega,{\bf q})=  \sum_{m_t} \int dE d{\bf p}
P^{m_t}(E,{\bf p})w^{m_t}_{\mu\nu}(p,q), \label{2} \ee
%\bigskip
of the nucleon response $w^{m_t}_{\mu\nu}(p,q)$ ($m_t=1/2$ and
$-1/2$ for  protons and neutrons respectively)  with the nuclear
spectral function $P^{m_t}(E,{\bf p})$. This gives the joint
probability of finding a nucleon with three momentum $\pb$ inside
the target nucleus, and remove it with an energy $E = E_B +
E_{A-1}^{\mbox{\scriptsize exc}}$. $E_B=M_{A-1}+M-M_A$ is the
nucleon binding energy and $E_{A-1}^{\mbox{\scriptsize exc}}$ the
excitation energy in which the residual nucleus is left. Notice
that for an off-shell nucleon, the energy $p_0=p_0(E,{\bf p})$
depends on its removing energy and  its three-momentum, thus to
implement the PWIA or any extension including FSI one  must
address some important questions. First, the nucleon structure
function is determined experimentally from proton or deuteron
scattering on  on-shell (free) targets, being $p^2=M^2$ (or $p_0 =
E_\pb \equiv \sqrt{\pb^2+M^2}$). In our case  we treat with an
off-shell bounded nucleon with $p_0 \neq E_\pb $, and
$p_0=p_0(E,\pb)$ depends on  how the binding effects are included.
Second, we need to extend the on-shell nucleon structure function
to the off-shell regime to use it as input in the nuclear response
calculation. The minimal hypothesis adopted in majority of works
is to assume that $w^{m_t\mbox{\scriptsize
(off-shell)}}_{\mu\nu}(p,q) = w^{m_t\mbox{\scriptsize
(on-shell)}}_{\mu\nu}(\widetilde{p},\widetilde{q})$, where
$\widetilde{p}$ and $\widetilde{q}$ depend on the off-shell
prescription  adopted for $p_0=p_0(E,\pb)$ . Third, whatever is
the $(\widetilde{p},\widetilde{q})$ pair  we have a lack of the
electromagnetic gauge invariance because $w^{m_t\mbox{\scriptsize
(off-shell)}}_{\mu\nu}q^\nu \neq 0$, due to the on-shell to
off-shell extension.  This brings in additional complications,
 a procedure being required to restore current conservation
\cite{Cio91,Kim94}.

The old data coming from the NE3 SLAC experiment \cite{Day87} were
analyzed within different approaches. One of the first PWIA
calculations  included  one-hole (1h) and two-particle - one-hole
(2p-1h) excitations in $P^{m_t}(E,{\bf p})$ \cite{Cio91,Bodek} .
The cross sections, when expressed in terms of the well known $x
\equiv Q^{2}/(2M \omega)$ Bjorken variable,  have been fairly well
reproduced in the quasielastic peak \footnote{The quasielastic
peak energy for a nucleon at rest corresponds to
$\omega_{qe}\equiv Q^2/2M$, which for ${\rm q}/M\gg 1$ leads to
$\omega_{qe} \simeq {\rm q}$.} ($x \simeq 1$) and inelastic
($x<1$)  regions, but underestimated for $x >1$.
In order to correct this the FSI were introduced in different ways .
For instance, when the PWIA  has been extended by
assuming a factorization hypotesis for the final nucleus wave function and by introducing pair
correlations \cite{Cio94}, the discrepancies in the region $1<x<2$ where circumvented.
For $x>2$ more than two nucleons should be
involved in the scattering process and thus the use of an optical
potential was required \cite{Cio96}.
%However, the effect of
%combining this potential with pair correlations has been analyzed
%for the scaling function but not for the cross section in
%different $\theta$ final electron geometries.
Benhar et al. \cite{Benhar91} have improved the PWIA results by
introducing the FSI through an optical potential and by generating
a folding function from the  multiple-scattering Glauber theory
and  color transparency.\linebreak The quite  recent CEBAF $4.05
GeV$ electron scattering \cite{Arr99} experiment covers the range
$1<Q^2<7 (GeV/c)^2$ and $0.2 \lsim x \lsim 4.2$, and vastly extend
the angular and energy-loss range of the older NE3 SLAC one. Rinat
and Taragin \cite{Gur95,Rin97} analyzed these results adopting an
alternative approach to the PWIA. The nuclear response function is
treated in a relativistic extension of the Gersch-Rodriguez-Smith
series \cite{Ger73}, while the FSI were introduced through binary
collisions. The CEBAF data  are well reproduced for $x<1$ and in
the left hand side neighborhood of the quasielastic peak  $x \gsim
1$ \cite{Rin00}. Nevertheless, for all $\theta$ geometries the
calculated cross section overestimates the data by a factor up to
2-10 in the low energy lost region ($x > 1$), being these
discrepancies associated to defects in the adopted  momentum
distribution. In the present work we develop a modified  version of the
PWIA (in the sense that the nucleon behaves as free one but with an effective mass)
where the off-shell effects and FSI are included via the RHA,
being at the same time the gauge invariance preserved. The Fermi
smearing effects are incorporated through a new  nucleon momentum
distribution, obtained from a perturbative calculation in nuclear
matter. In addition, different parameterizations for the inelastic
nucleon response measured at SLAC, are analyzed. The CEBAF data
are  satisfactorily reproduced for all measured
geometries, taking into account that the cross section varies over many orders
of magnitude, the mentioned overestimation being avoided  in the $x >
1$ region.

\section{Elastic and inelastic cross sections}

As the electron probes a region of dimensions $1/{\rm q}$, for
high momentum transfers, surface effects are supposed to be of minor
importance and the nuclear matter framework is adopted. How good is this
assumption will be analyzed in Section IV, where the
theoretical  results will be compared  with the experimental
and nuclear matter extrapolated data.
The
nucleon will be bounded by interaction with the scalar $\phi$ and
vector $V_\mu$ mesons fields, within the framework of quantum
hadrodynamics(QHDI)\cite{Serot86,Ferre}. The nucleus response
tensor is the Lorentz invariant amplitude and reads \cite{Bjorken}

\br W^{A}_{\mu\nu}(\omega,{\bf q})&=& {{\rm k}M_A\over
\sqrt{(k.P_A)^2 - m^2M_A^2}} {V  \over (2\pi)^3} \nonumber\\
&\times& \sum_{\pb'm_s'm_t'} \sum_{f} \bra{P_A} \hat{J}(0)_\mu
\ket{p'm_s'm_t',P_{A-1}^f}\nonumber\\
&&\bra{p'm_s'm_t',P_{A-1}^f}\hat{J}(0)_\nu \ket{P_A} \nonumber \\
&\times& (2\pi)^4 \delta(P_A + k - k'-P_{A-1}^f-p'),\label{A7} \er
being  $P_{A}=(M_A,0)$ and $P_{A-1}^f= (\sqrt{\pb_f^2 +
(M_{A-1}^{f})^2},\pb_f)$ the target and residual nucleus
four-momentum respectively, with  the mass $M_{A-1}^{f}=M_{A-1} +
\omega_{A-1}^f$ including the excitation energy $\omega_{A-1}^f$.
The sum on $f$ encloses the set of final residual nucleus states.
We also sum on the final states of the struck nucleon with
four-momentum $p'=(p'_0,\pb')$, spin $m_s'$ and isospin $m_t'$,
with density $V/(2\pi)^3$ in the quantization volume $V$.
$\hat{J}(x)$ is the effective hadron  current density operator
$\hat{J}_\mu(x) = i \psixb \Gamma_\mu(q)\psix$ with
$\Gamma_\mu(q)= F_1(q^2) \gmu + i F_2(q^2) {\kappa \over 2M}\smunu
q^\nu$ for the nucleon elastic response case, being $\psix$ and
$\kappa$ the nucleon field and anomalous magnetic moment,
respectively.

We are going develop on the same footing the nuclear
response calculation within the mean field theory (MFT) (where the
meson fields are approximated  by their vacuum spectation, i.e.
constant, values), and in the RHA\cite{Serot86} (where vacuum
fluctuation corrections are added to the MFT results). Later, when
we  compare the calculated cross section with the data, the RHA
election will be  justified.
The nucleon field is expanded as
\br \psix &=& {1\over\sqrt{V}} \sum_{\pb m_s m_t} \sqrt{M^* \over
E^*_\pb} \left[ \u{\pb m_s m_t} a_{\pb m_s m_t} e^{ip \cdot
x}\right. \nonumber \\ &+& \left. b^\dag _{\pb m_s m_t} \v{\pb m_s m_t}
e^{-ip \cdot x}\right],\label{A9} \er where the single particle
spectrum is given by
\br p_0 = C_{V}^{2} {\rho_{B} \over M^2}+ E^*_\pb,\label{7} \er
\begin{figure}[h!]
%\vspace{-1.cm}
  \begin{center}
    \includegraphics[height=8.cm,width=6.0cm]{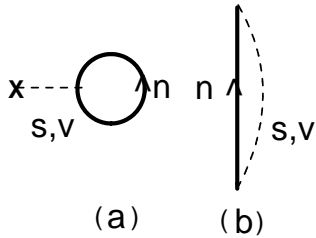}
  \end{center}
\vspace{-4.cm} \label{tad} \caption{(a) Tadpole diagram included in the MFT and RHA
self-energies. (b) Tadpole exchange diagram that is added in order to get the relativistic
Hartree Fock self-energy. The dashed lines indicate the propagator  of the scalar (S) or vector
meson (V) that interacts with a nucleon n (full lines) }
\end{figure}
with $E^*_\pb=\sqrt{{\bf p}^{2} + M^{*2}}$ and $M^{*}\equiv M +
\Sigma(C_{S},M^{*})$. $M^*<M$ is the effective mass acquired by
the nucleon by action of the attractive scalar field and is
determined self-consistently \cite{Serot86} through the scalar
self-energy $\Sigma \equiv \Sigma_{MFT}$ or $\Sigma_{RHA}$.
$\Sigma_{MFT}$ includes the tadpole diagram a) in
the Figure 1, retaining in its evaluation only the contribution from nucleons in the
filled Fermi sea in the nucleon propagator (tick full lines). $\Sigma_{RHA}$
includes the same diagram but the full nucleon propagator (which encloses the contribution
of the occupied negative-energy states) is used
in the evaluation of the self-energy. Then
the MFT or the RHA are derived by summing up the self-energy to all orders
through the self-consistent determination of $M^*$, being
this procedure  convergent in both cases.
The first term in Eq.\rf{7} accounts for  the action of the repulsive
vector field. $C_{V}$ and $C_{S}$ are the two free parameters
\cite{Walecka}, which depend on the meson coupling constants and
masses, fixed to reproduce  the experimental binding energy per
nucleon of $-16$ MeV  at the Fermi momentum $p_F = 1.42 fm^{-1}$
(or  the baryon density $\rho_{B}=0.19 fm^{-3}$) for  the  normal
nuclear matter.
Assuming that the residual nucleus is left in its
ground state and adopting the prescriptions ii) and iii) mentioned
above, the response tensor  can be obtained from Eqs. \rf{A7} to
\rf{7} as

\br
W^{A}_{\mu\nu}(q)&=& 2 \sum_{m_t} \int d\pb { M^*\over
E^*_\pb}n^{m_t}(\pb)w^{m_t}_{\mu\nu}(p^*,q),\label{8} \er
where the factor 2 resembles the sum over spin states, and

\br
w^{m_t}_{\mu\nu}(p^*,q)&=& w_{e 1}^{m_t}(Q^2,\nu^*)[-g_{\mu\nu}+\frac{ q_\mu q_\nu}{q^2}]
\nonumber\\&+&
w_{e 2}^{m_t}(Q^2,\nu^*) [\frac{ p^{*}_\mu}{M^*}-\nu^*\frac{q_{\mu}}{q^{2}}]
[\frac{p^{*}_\nu}{M^*}-\nu^*\frac{q_{\nu}}{q^{2}}],\nonumber\\\label{9} \er
with  $p^* = (E^*_\pb,\pb)$ and \footnote{$Q^2$ and $\nu =
p.q / M$ are commonly used as independent variables for $w_{1,2}^{m_t}$ in the nucleon response.}
$\nu^*=p^*\cdot
q/M^*$ . $n^{m_t}(\pb)$ is the  nucleon momentum distribution in the
target ground state $\ket{0_A}$

\be n^{m_t}({\bf p})= {V \over (2\pi)^3}\bra{0_A}a^\dag_{\pb m_s
m_t}a_{\pb m_s m_t}\ket{0_A},\label{A14} \ee
 normalized as $2 \int d\pb \ n^{m_t}({\bf p}) =
N^{m_t}$, with $N^{m_t} = Z , N$ for $m_t=1/2,-1/2$. The elastic
Lorentz scalar functions present in \rf{9} are

\begin{equation}
w_{e 1}^{m_t}(Q^{2},\nu^*)= \tau
G_{M}^{m_t 2}(Q^{2})\delta(\nu^* - \frac{Q^{2}}{2M^*})\label{A17}
\end{equation}
\begin{equation}
w_{e 2}^{m_t}(Q^{2},\nu^*)= \frac{G_{E}^{m_t 2}(Q^{2})+\tau
G_{M}^{m_t 2} (Q^{2})}{1+\tau}\delta (\nu^* -
\frac{Q^{2}}{2M^*})\label{A18}
\end{equation}
where  $G_{E}^{m_t}(Q^{2})=F_{1}^{m_t}(Q^2) - F_{2}^{m_t}(Q^2)
\kappa^{m_t}\tau$ and $G_{M}^{m_t}(Q^{2})=F_{1}^{m_t}(Q^2) +
F_{2}^{m_t}(Q^2)\kappa^{m_t}$ are the electric and magnetic form
factors, and $\tau = Q^{2}/4M^{*2}$. In the numerical calculations
we adopt the Sachs form for them, assuming  that they do not
change in the nuclear medium \cite{Koltun}. Equations \rf{9},
\rf{A17} and \rf{A18}  show that  the MFT or  RHA lead to the
prescription $w^{m_t\mbox{\scriptsize (off-shell)}}_{\mu\nu}(p,q)
= w^{m_t\mbox{\scriptsize (on-shell)}}_{\mu\nu}(p^*,q)$  and
$w^{\mbox{\scriptsize (off-shell)}}_{e \ 1,2}(Q^2,\nu) =
w^{\mbox{\scriptsize (on-shell)}}_{e \ 1,2}(Q^2,\nu^*)$, for the
elastic case. The nucleon spinors carry  a four momentum $p^*$
being  $p^{*2}=M^{*2}$, and as $M^*<M$ this makes us remember  that the
struck nucleon is bounded. Lorentz , parity  and  gauge
invariances are now also fulfilled as were for a nucleon of mass
$M$, as consequence of the form of the Eq.\rf{9} \cite{Drell}. FSI
are included, since the nucleon is bounded also after the
interaction with the photon. For $Q^2>1$ (GeV/c)$^2$ the
probability of exciting internal states of the nucleon is
important, and a replacement  $w_{e \ 1,2}^{m_t} \rightarrow w_{
1,2}^{m_t}=w_{e 1,2}^{m_t} + w_{i 1,2}^{m_t}$ in \rf{9} should be
done, adding an inelastic  contribution $w_{i 1,2}^{m_t}$.
For $w^{m_t}_{i 1,2}$
we use different parametric fits done at SLAC for $p(e,e')p'$ and
$d(e,e')d'$ data through  the Eqs.\rf{9},
with $M^*=M$. We assume that the recipe
$w^{{off-shell}}_{i1,2}(Q^2,\nu) =
w^{{on-shell}}_{i1,2}(Q^2,\nu^*)$, which naturally appeares in the
elastic case, is also valid for the inelastic  nucleon response
function. Finally, the decomposition $w_{1,2}^{m_t}=w_{e
1,2}^{m_t} + w_{i 1,2}^{m_t}$ leads  also to split the  inclusive
cross section \rf{1} in elastic and inelastic contributions.

\section{Nucleon momentum distribution}

From Eqs. \rf{2} and \rf{8} it is clear that we are working within the FSA
 $P^{m_t}(E,{\bf p}) \sim
n^{m_t}(\pb) \delta(E-E_B)$, giving  $n^{m_t}(\pb)$
the probability of finding a nucleon with momentum $\pb$, and isospin
$m_t$ in the target $\ket{0_A}$. $n^{m_t}({\bf p})$ is
calculated
in a 0p0h + 2p2h + 4p4h configuration space for the A-target,
being
\br &&|0_A\rangle  =  {\cal N} \left[|0p0h \rangle +
\frac{1}{(2!)^2} \sum_{p's,h's} c_{p_1p_2h_1h_2}
|p_1p_2h_1h_2\rangle \right .\nonumber \\ & + & \left
.\frac{1}{(4!)^2} \sum_{p's,h's} c_{p_1p_2p_3p_4h_1h_2h_3h_4}
|p_1p_2p_3p_4h_1h_2h_3h_4\rangle \right  ],\nonumber\\ \label{10b}
\end{eqnarray}
where these $|npnh\rangle$, (with $n=0,2,4$) stand for the unperturbed states.
In this way in the residual nucleus
we have 1h, 2p3h, 4p5h, 1p2h,and 3p4h excitations  when the struck nucleon is removed.
The residual nucleon-nucleon interaction is included  within a perturbative
approach as in Ref.\cite{Mar96} by expanding the coefficients $c_{2p2h}$  and $c_{4p4h}$ up to
the first and second order, respectively.
This ''minimum''
perturbative scheme allows  to include norm corrections
${\N} = \langle 0_A|0_A \rangle^{-1}$, avoiding in this way contributions of
unbalanced disconnected diagrams. We get

\br n^{m_t
}({\bf p})&=& {3N^{m_t}
 \over 4 \pi {\rm p}_F^3} \left[
\theta(1-{\rm p})+ \delta n^{(2)}({\rm p}) + \delta n^{(4C)}({\rm
p})\right],\nonumber \\ \label{10} \er where ${\rm p}\equiv |\pb|$
is measured in units of the Fermi momentum $p_F$. The first term
is the usual 0p0h Fermi step function, while $\delta n^{(2)}({\rm
p})$ and $\delta n^{(4C)}({\rm p})$ (where the superscript $C$
indicates ''connected'' 4p4h diagrams) enclose  2p2h and 4p4h
contributions respectively, which deplete it. The expressions for
$\delta n^{(2)}({\rm p})$ and $\delta n^{(4C)}({\rm p})$, are
given in Ref. \cite{Mar96}, while the Goldstone diagrams
corresponding to them  are shown in Figure 2.
\begin{figure}
\vspace{5.5cm}
\hspace{-9cm}
\begin{picture}(50,75)(15,25)
\setlength{\unitlength}{0.75pt} \thicklines \put (200,300){\oval
(30,70)} \put (200,300){\oval (16,70)} \put (200,335){\circle*{6}}
\put (200,265){\circle*{6}} \put (185,300){\circle{6}} \put
(185,300){\circle*{3}} \put (190,235){{\bf \large (a)}} \multiput
(100,150)(80,0){4}{\oval (30,50)[r]} \multiput
(100,150)(80,0){4}{\oval (16,50)[r]} \multiput
(100,175)(80,0){3}{\line (0,-1){50}} \multiput
(100,175)(80,0){4}{\circle*{6}} \multiput
(100,125)(80,0){4}{\circle*{6}} \multiput (70,185)(80,0){3}{\line
(1,-2){30}} \multiput (70,115)(80,0){3}{\line (1, 2){30}}
\multiput (70,185)(80,0){3}{\line (0,-1){70}} \multiput
(70,150)(80,0){4}{\oval (30,70)[l]} \multiput
(70,150)(80,0){4}{\oval (16,70)[l]} \multiput
(70,185)(80,0){4}{\circle*{6}} \multiput
(70,115)(80,0){4}{\circle*{6}} \put (55,150){\circle{6}} \put
(55,150){\circle*{3}} \put (150,150){\circle{6}} \put
(150,150){\circle*{3}} \put (242,160){\circle{6}} \put
(242,160){\circle*{3}} \put (75,90){{\bf\large (b)}} \put
(155,90){{\bf\large (c)}} \put (235,90){{\bf\large (d)}} \put
(315,90){{\bf\large (e)}}
%\put (180,35){{\bf\large Figure 1}}
\put (295,150){\circle{6}}
\put (295,150){\circle*{3}}
\put (308,185){\line (1,-2){30}}
\put (308,115){\line (1, 2){30}}
\put (312,185){\line (1,-2){30}}
\put (312,115){\line (1, 2){30}}
\end{picture}
\vspace{-1cm}
\caption{Goldstone diagrams corresponding to the second-order 2p2h correction
$\delta n^{(2)}(\rm p)$ (a) and fourth order 4p4h correction
$\delta n^{(4C)}(\rm p)$ (b,c,d,e). Each line indicates schematically a particle
or a hole state, the dots represent the residual interaction and
the encircled dots correspond to the number operator $n(\pb)$.
With b, c, d and e
we indicate different ways to attach the number operator to a particle or hole line
in $\delta n^{(4C)}(\pb)$.}
\end{figure}
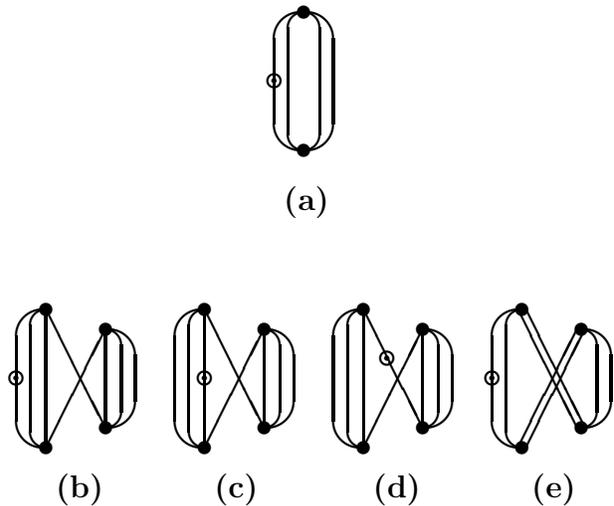

\section{Results and conclusions}
\label{results}
We now compare the differential  cross section
 calculated within our model, with the CEBAF
experimental results for $^{56}$Fe \cite{Arr99} for the various
accessible geometries \linebreak $\theta = 15, 23, 30, 37, 45, 55, 74^0$.
One of the parameterizations for $w^{m_t}_{i 1,2}$
was found by Bodek et al. \cite{Bodek} in the kinematical range $1<Q^2<20$
(GeV/c)$^2$ and $0.1 \leq x \leq 0.77$.
 The other one was reported by Whitlow
\cite{Whitlow}, and corresponds to
the range $0.6<Q^2<30$ (GeV/c)$^2$ and $0.06 \leq x \leq 0.9$.

\begin{figure}[h!]
  \begin{center}
    \includegraphics[height=12.cm,width=10.0cm]{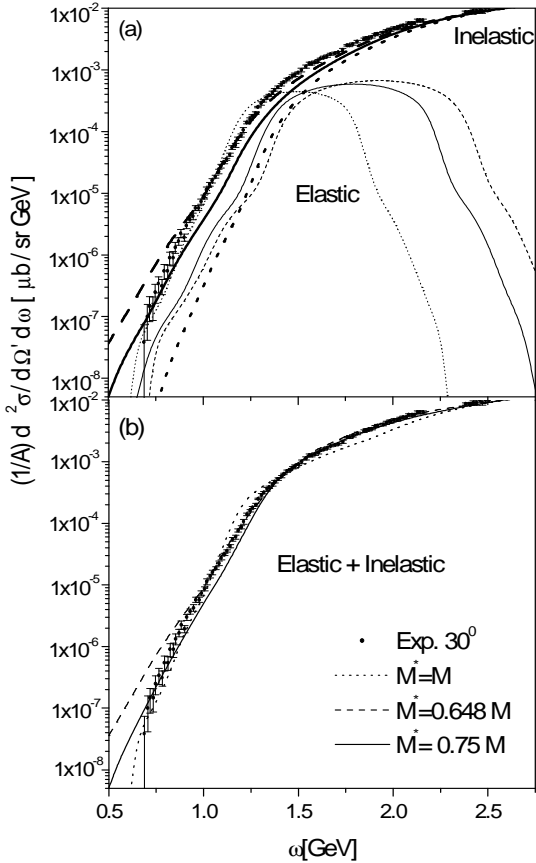}
  \end{center}
  \vspace{-1cm}
\caption{ We show the sensibility with the effective mass $M^*$ of
the quasielastic and inelastic contributions to the cross sections
per nucleon for $^{56}Fe$. Here the replacement
$\epsilon'=\epsilon + \omega$ is done. In the panel (a) both cross
sections are shown separately for the values $M^*= 1$, $0.64$, and
$ 0.74$. Thin lines indicate elastic cross sections while thick
lines indicate the inelastic one. In panel (b) the total elastic +
inelastic cross section is shown for the different values of
$M^*$. Again, experimental results come from Ref. \cite{Arr99}.}
\end{figure}

The functions obtained in these parameterizations are described in
detail in Ref. \cite{Ferre}, and as they do not cover all the low
energy and momentum transfer region of  CEBAF, an extrapolation is
necessary. This fact could introduce some uncertainties in the
calculation. Within the MFT  and for $^{56}Fe$ ($p_F=1.36$
fm$^{-1}$), $M^*=0.648 M$. This  value is too low to reproduce
satisfactorily the total cross section since the quasielatic peak
is shifted  too much to  the right and its width ($\Delta
\omega_{qe}$) is enlarged in excess, as shown in the \nolinebreak Figure 3.
This behavior of the MFT at high momentum transfers was analyzed
in Ref.\cite{Kim94} for the longitudinal response, where a
dependence of $\phi$ and $V_\mu$ with ${\rm p}$ was introduced
\footnote{Within the Hartree-Fock approach this dependence is generated
naturally by adding in the self energy the exchange graph b) of the Figure 1 , and summing up
to all orders.}. The value of $\omega_{qe}$ is certainly corrected
(but not the detailed peak's shape)
since as it is shown, the binding-energy shift effects are
diminished as ${\rm p}$ increases. Nevertheless the ${\rm p}$
dependence of the fields carries  a gauge invariance violation
that is corrected by introducing a (nonunique) vertex correction
in $\Gamma_\mu$. The Eqs.\rf{9}, \rf{A17} and \rf{A18} are
altered, and the prescription $w^{{off-shell}}_{e 1,2}(Q^2,\nu) =
w^{{on-shell}}_{e 1,2}(Q^2,\nu^*)$ is no longer  valid. This
brings a problem at the moment of introducing the inelastic
response, since the assumption $w^{{off-shell}}_{i 1,2}(Q^2,\nu) =
w^{{on-shell}}_{i 1,2}(Q^2,\nu^*)$, is not justified.
Alternatively,  we try to improve the MFT  description by adding
the vacuum fluctuation corrections to $\Sigma_{MFT}$
\cite{Serot86}, and go to the RHA \cite{Chin} where $M^*=0.74M$.
The Eqs. \rf{9}, \rf{A17} and \rf{A18}, are now still valid. As
can be seen in Figure 3 the RHA the binding-energy shift is more
moderated and the width is diminished, getting a better
description for  the total cross section. This improvement is not
casual since as it is well known the RHA yields to the ''best''
single-particle spectrum in the sense that it minimizes the energy
of the whole system. In addition when the longitudinal response
for electron scattering at ${\rm q}=0.55 GeV$ and $1.14 GeV$
transfers is analyzed, we get a value $\omega_{qe}=0.182 GeV$
($\Delta \omega_{qe}=0.302 GeV$) and $0.615 GeV$ ($0.436 GeV$)
respectively, in fully agreement with the results reported in
Ref.\cite{Kim94}. This indicates that keeping ${\rm p}$-independent fields and
changing the value of $M^*$ in Eqs. \rf{A17} and \rf{A18},
one can still improve the quasielastic peak position both at low and high momentum transfers
at the same time. FSI are taken into account in our model at the
RHA level. Binding effects are present in the final state, since
the nucleon still has mass $M^*$ after absorbing the photon. This
simple form of introducing FSI has never been  used previously to
describe a multi-GeV electron experiment with the inclusion of the
inelastic nucleon response, being only described  in the past  the
quasielastic cross section at intermediate energies in the MFT
framework \cite{Rosenf}.

\begin{figure}[h!]
  \begin{center}
    \includegraphics[height=14.cm,width=10.0cm]{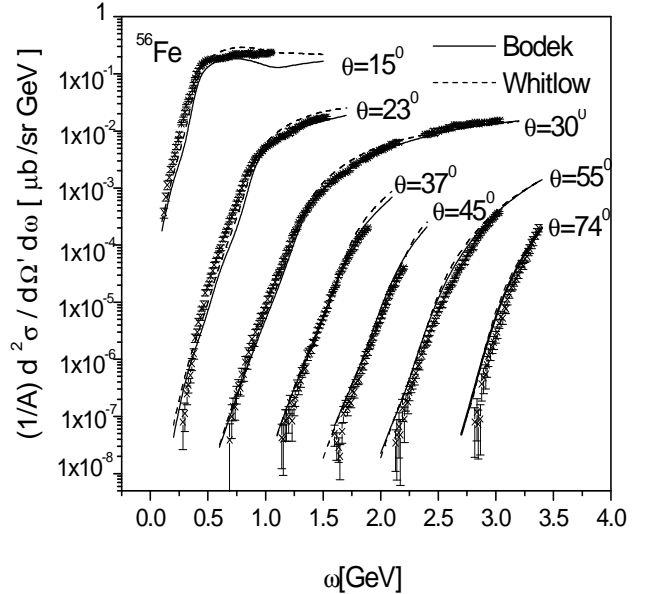}
  \end{center}
\vspace{-6cm}
 \caption{ Calculated differential cross section per
nucleon for different $\theta$ geometries for $^{56}$Fe.
Experimental data come from Ref. \cite{Arr99}. Results are shown
for both, the fitting of Bodek and Withlow of the inelastic
nucleon response, and for a value $M^*=0.74$ corresponding to the
RHA.}
\end{figure}
FSI affect directly the quasielastic response defined in Eqs.
\rf{9}, \rf{A17} and \rf{A18}, since the size and position of the
quasielastic peak are controlled by $\tau = Q^{2}/4M^{*2}$ (which
scales quadratically with $M^*$) and $\nu^*$, respectively. In the
inelastic nucleon response, FSI effects are included indirectly
through the replacement $\nu \rightarrow \nu^*=p^*.q/M^*$ in the
$w_{i 1,2}^{m_t}$ on-shell functions. Our results for the total
cross section are shown in Figure 4. As  can be seen in,
the overall agreement is good for all angles $\theta$,
 considering that that the cross section varies over several decades.
At $\omega < \omega_{qe}$ ($x>1$) Withlow's fit seems to be
preferred to Bodek's, which is due possibly
 to differences in the extrapolation
for the $x>1$ range. For $\omega > \omega_{qe}$ ($x<1$) the
behavior is opposite.  We see that the model tends to overestimate
the $x>1$ data,  in the last two $\theta$ values. The inelastic
response dominates the cross section at these geometries since
$Q^2 \gsim 4 (GeV/c)^2$, and this overestimation  could  be also
as consequence of uncertainties in the extrapolation  for $x>1$.
We conclude  that to implement the  generated nuclear matter
momentum distribution within the RHA framework, is a consistent
approach for treating electron scattering at these momentum
transfers. This is further supported by the following
observations:

i)In the Figure 5 we compare the nuclear
matter extrapolated cross section reported in Ref.\cite{Day87} for
$\epsilon= 4 GeV$ and $\theta = 30^0$ in the shown energy lost
region, with the experimental results of Ref.\cite{Arr99} and our
calculations. As can be seen the extrapolated results for this geometry
are larger than the experimental data, being the difference small comparatively
to the range of variation of the cross section (in the Figure 5 we show only
an interval of the full energy lost region). This
indicates that nuclear matter is a reasonable framework at
these momentum transfers. Our nuclear matter results
change from below the data to above the extrapolation, indicating
that some improvements as  would be a momentum dependence of the effective mass or
higher-order contributions to the momentum distribution should be
more deeply analyzed;
\begin{figure}[h!]
\vspace{-0.cm}
  \begin{center}
    \includegraphics[height=10.cm,width=8.0cm]{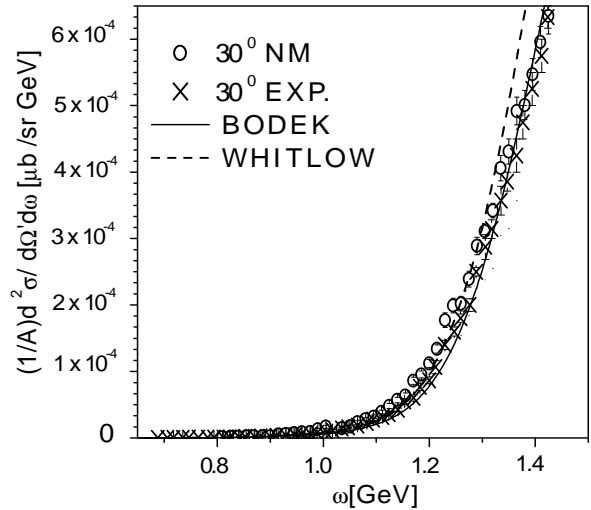}
  \end{center}
\vspace{-2.5cm} \label{comparation} \caption{Comparison of the
extrapolated nuclear matter (NM) cross section per nucleon at
$\epsilon =4. GeV$ reported in Ref.\cite{Day87}, with the
experimental data at $\epsilon =4.05 GeV$ \cite{Arr99} and our
calculations, for the indicated geometry}
\end{figure}

ii)We are working within the Fermi smearing approach that
in previous calculations with other nuclear matter momentum
distributions \cite{Fan84,Ben89}, lead to a large overestimation
of the cross section at low electron energy loss
\cite{Cio91,Ben94}. As can be seen from Eq.\rf{10} and
Figure 2, we are including fourth-order corrections to $n(\pb)$ in
addition to the usual second-order contributions \cite{Fan84}. In
Ref.\cite{Mar96}, it has been shown that the second-order
perturbation approach overestimates the depletion of the Fermi
surface and thus the high momentum tail of the momentum
distribution. This could be the reason of the mentioned
overestimation present in  previous nuclear matter calculations. In the
Figure 6 we show the momentum distribution obtained from the
Eq.\rf{10} together with its second order approach, being $\delta
n^{(4C)}({\rm p})$ dropped. In the same figure we show the
momentum distribution of Ref.\cite{Fan84} (parameterized in
\cite{Cio96}), which was obtained within a second order
perturbation approach over a  set of unperturbed variational wave
functions.
\begin{figure}[h!]
\vspace{0.5cm}
  \begin{center}
    \includegraphics[height=9.cm,width=9.0cm]{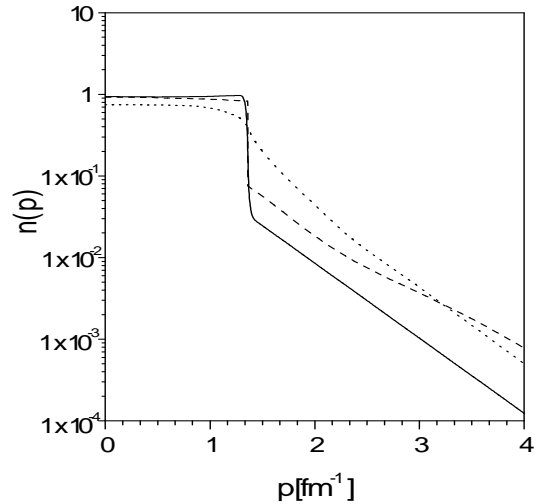}
  \end{center}
\vspace{-2.cm} \label{npdis} \caption{Comparison of the nuclear
matter momentum distribution used in our calculations (full
lines), with its second order approach (dotted lines) and the
momentum distribution of Ref.\cite{Fan84} (dashed lines).}
\end{figure}
It is important to mention that FSI also are responsible for the
behavior of the cross section in the low energy lost region. In
our model FSI are taken into account by using an effective nucleon
mass, and different values of $M^*$ lead to different
contributions to the cross section in the mentioned region, as can
be seen from Figure 3. The value of $M^*=0.74 M$ within the RHA,
seems to introduce FSI consistently with the implemented momentum
distribution;
\begin{figure}[h!]
%\vspace{-0.75cm}
  \begin{center}
    \includegraphics[height=14.cm,width=10.0cm]{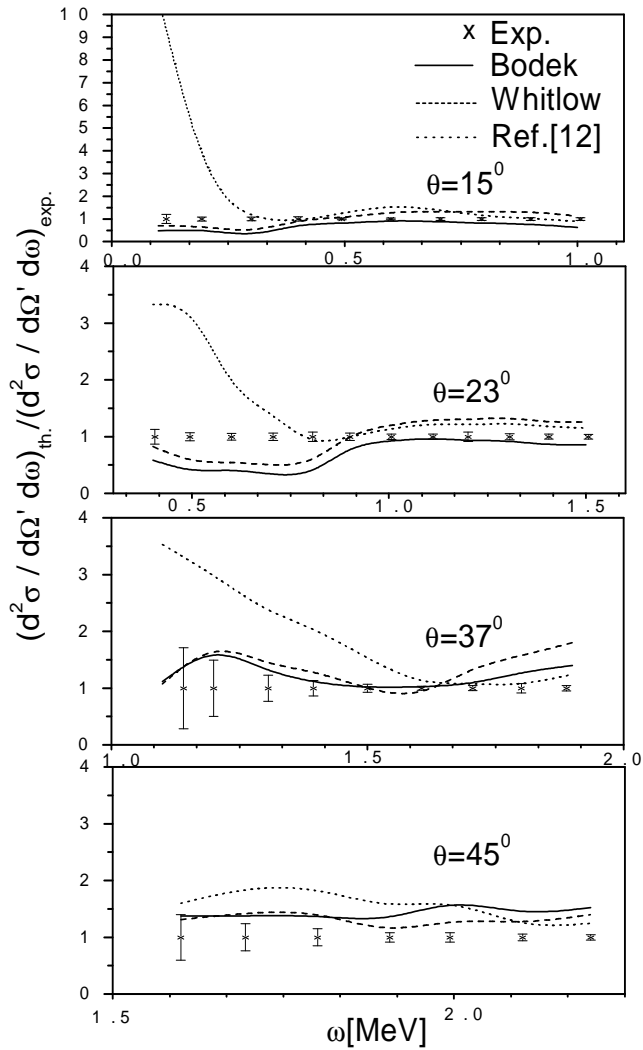}
  \end{center}
\label{comparation} \caption{Ratio of the theoretical to
experimental cross section for some selected geometries, within
our model (full and dashed lines) and the model of
Ref.\cite{Rin00} (dotted lines). }
%\vspace{-1.cm}
\end{figure}

iii)Finally the  overestimation by
a factor 2-10 in previous theoretical evaluations of the cross
section at $x>1$ \cite{Rin00}, is not present in our calculation.
This can be seen in Figure 7 where we compare the ratios of the
theoretical to experimental cross section, for some selected
geometries where the differences are appreciable, in our model and
in that of Ref.\cite{Rin00}.

In summary, to treat the scattering of GeV electrons by nuclei we
have implemented a new Fermi smearing approach. Binding
effects and FSI are introduced through the nucleon effective mass
within the RHA, that leads to better results than the plain MFT
\cite{Ferre}. In the model, current conservation is preserved
naturally without ad-hoc modifications in the structure functions.
Fermi smearing effects are introduced through a new momentum
distribution   that accounts for 2p2h and 4p4h correlations in the
target, generated  via a perturbative approach in nuclear matter.
We get a reasonable overall description of the behavior of the measured cross section
at CEBAF, for the scattering of $4.05$ GeV electrons on
$^{56}$Fe. The agreement for  all the accessible
geometries, has been significantly improved in
comparison with previous theoretical studies \cite{Rin00}. It
could also suggest that within the model the FSI and Fermi
smearing effects combine consistently, this could be more clearly
established examining the effect of changing  $M^*$ on the
contributions to the cross section coming from the hole and
particle strength functions. This more detailed analysis and the
scaling behavior of the model will be reported elsewhere
\cite{Mariano03}.

Acknowledgements: The Work of A. M. was partially supported by Conicet (Argentina).

\end{document}